# Magnetostrictive thin films for microwave spintronics.

D.E. Parkes[1†], L.R. Shelford[2†], P. Wadley[1,3], V. Holý[3], M. Wang[1], A.T. Hindmarch[1a)], G. van der Laan[2], R.P. Campion[1], K.W. Edmonds[1], S.A. Cavill[2*] and A.W. Rushforth[1*].

[1]*School of Physics and Astronomy, University of Nottingham, Nottingham NG7 2RD, United Kingdom.*

[2]*Diamond Light Source, Chilton, Didcot, Oxfordshire OX11 0DE UK.*

[3]*Faculty of Mathematics and Physics, Charles University in Prague, Ke Karlovu 3, 121 16 Prague 2, Czech Republic.*

a) Present address: Centre for Materials Physics, Department of Physics, Durham University, South Road, Durham DH1 3LE, United Kingdom.

†These authors made equal contributions to the work.

[*] Address to whom correspondence should be sent.

Stuart.Cavill@diamond.ac.uk, Andrew.Rushforth@nottingham.ac.uk

*Multiferroic composite materials, consisting of coupled ferromagnetic and piezoelectric phases, are of great importance in the drive towards creating faster, smaller and more energy efficient devices for information and communications technologies. Such devices require thin ferromagnetic films with large magnetostriction and narrow microwave resonance linewidths. Both properties are often degraded, compared to bulk materials, due to structural imperfections and interface effects in the thin films. We report the development of single crystal thin films of Galfenol ($Fe_{81}Ga_{19}$) with magnetostriction as large as the best reported values for bulk material. This allows the magnetic anisotropy and microwave resonant frequency to be tuned by voltage-induced strain, with a larger magnetoelectric response and a narrower linewidth than any previously reported Galfenol thin films. The combination of these properties make the single crystal thin films excellent candidates for developing tunable devices for magnetic information storage, processing and microwave communications.*

The ability to manipulate magnetic properties by the application of electric fields is an important functionality in designing devices for information storage and processing where it will remove the need to apply electrical currents which create heat dissipation and stray magnetic fields. Magnetoelectric effects are also important for creating electrostatically tunable microwave resonators, phase shifters and filters which find applications in signal processing technologies[1-3], and in schemes for performing logical processing operations using spin waves [4-6].

These effects can be realised by voltage-induced strain in ferromagnet/piezoelectric hybrid structures[7-13] containing a magnetostrictive ferromagnetic component, as was proposed recently in concepts for energy efficient magnetic random access memory (MRAM)[14] and spin wave logical processing devices[4,5]. The realisation of such nanoscale magnetoelectric devices requires the development of thin magnetic films with high magnetostriction constants. Galfenol, an alloy of Fe and Ga has attracted great interest over the last decade or so because it exhibits the largest magnetostriction of any metal alloy not containing a rare earth element[15], whilst retaining a high magnetic moment, large spin polarisation and large magnetic anisotropy. However, sputtered Galfenol in thin film form tends to show significant reductions of the magnetostriction due to the polycrystalline nature of the films[16]. Also, especially detrimental for microwave applications is the fact that the resonant linewidths are large[17,18]. To overcome this, FeGa can be doped with boron resulting in narrow linewidths. Unfortunately whilst alloying with boron reduces the linewidth it also reduces the magnetostriction further[19] and so the magnitude of the ME effect has to be traded off against linewidth. Narrow resonant linewidths and low damping are particularly important attributes of materials for microwave and spinwave applications.

In this article we report the development of single crystal thin films of $Fe_{81}Ga_{19}$ grown by molecular beam epitaxy (MBE) on GaAs(001) substrates. The close lattice match to the substrate allows the growth of single crystal films with a cubic structure. Previous reports of MBE grown Galfenol have not included measurements of the magnetostriction or of the microwave properties[20,21]. We have measured the magnetostriction by monitoring the change in the magnetic anisotropy in response to an applied strain (*i.e.* the Villari effect) and found it to be as large as the best values reported for bulk single crystals of Galfenol, and larger than for any other thin films not containing rare earth elements. Ferromagnetic resonance (FMR) studies reveal narrow resonant linewidths of 6.7 mT, and a strain tunable response of the resonant field of 19.3T per unit strain, larger than the best reported values for Galfenol based thin films.

**Results**

Structure

The single crystal structure of our MBE-grown thin film was confirmed by x-ray diffraction (XRD). Similar to the case of Fe/GaAs(001) [22], the (001) planes of the $Fe_{81}Ga_{19}$ film are parallel to those of the substrate, with a lattice constant which is approximately half that of GaAs. The 2θ/ω scan shown in figure 1(a) was fitted by standard XRD software using a model of a single perfect layer on a semi-infinite substrate. From the fit we determined the thickness of the layer to be t=21.0 ± 0.2 nm and the vertical strained lattice parameter $a_\perp$=(0.29607 ± 0.00003) nm, representing a lattice mismatch $(a_\perp - a_0)/a_0$, of 4.6%, where $a_\perp$ is the lattice parameter along the growth axis for the $Fe_{81}Ga_{19}$ film, and $a_0$=0.283nm corresponds to half the GaAs substrate lattice parameter. The presence of a sharp peak in the ω-scan (see inset to Fig. 1(a)) points towards the absence of extended defects within the film and the broad feature is likely caused by localised, point like defects.

Determination of the Magnetostriction

Measurement of the magnetostriction in a thin film is complicated by the fact that the film is clamped to a substrate, and must be achieved by measuring the change in the magnetic anisotropy in response to a strain induced in the film. To do this we employed a method developed previously in studying magnetostriction in dilute magnetic semiconductors[7,23]. This involved bonding the magnetic film to a piezoelectric transducer and inducing a uniaxial strain in the film by the application of a voltage to the transducer. The magnetic anisotropy was deduced by fitting to curves of magnetisation versus external magnetic field, which were extracted from electrical measurements on a Hall bar device fabricated in the $Fe_{81}Ga_{19}$ film. Electrical measurements allow readout of the direction of the magnetisation by measuring the anisotropic magnetoresistance (AMR). To a good approximation, for in plane magnetisation, the longitudinal ($\rho_{xx}$) and transverse ($\rho_{xy}$) resistivities are given by $\rho_{xx} = \rho_{av} + \Delta\rho\cos 2\theta$ and $\rho_{xy} = \Delta\rho\sin 2\theta$, where θ is the angle between the magnetisation and the current, $\rho_{av}$ is the average of $\rho_{xx}$ when the magnetisation is rotated through 360° in the plane, and Δρ is the amplitude of the AMR.

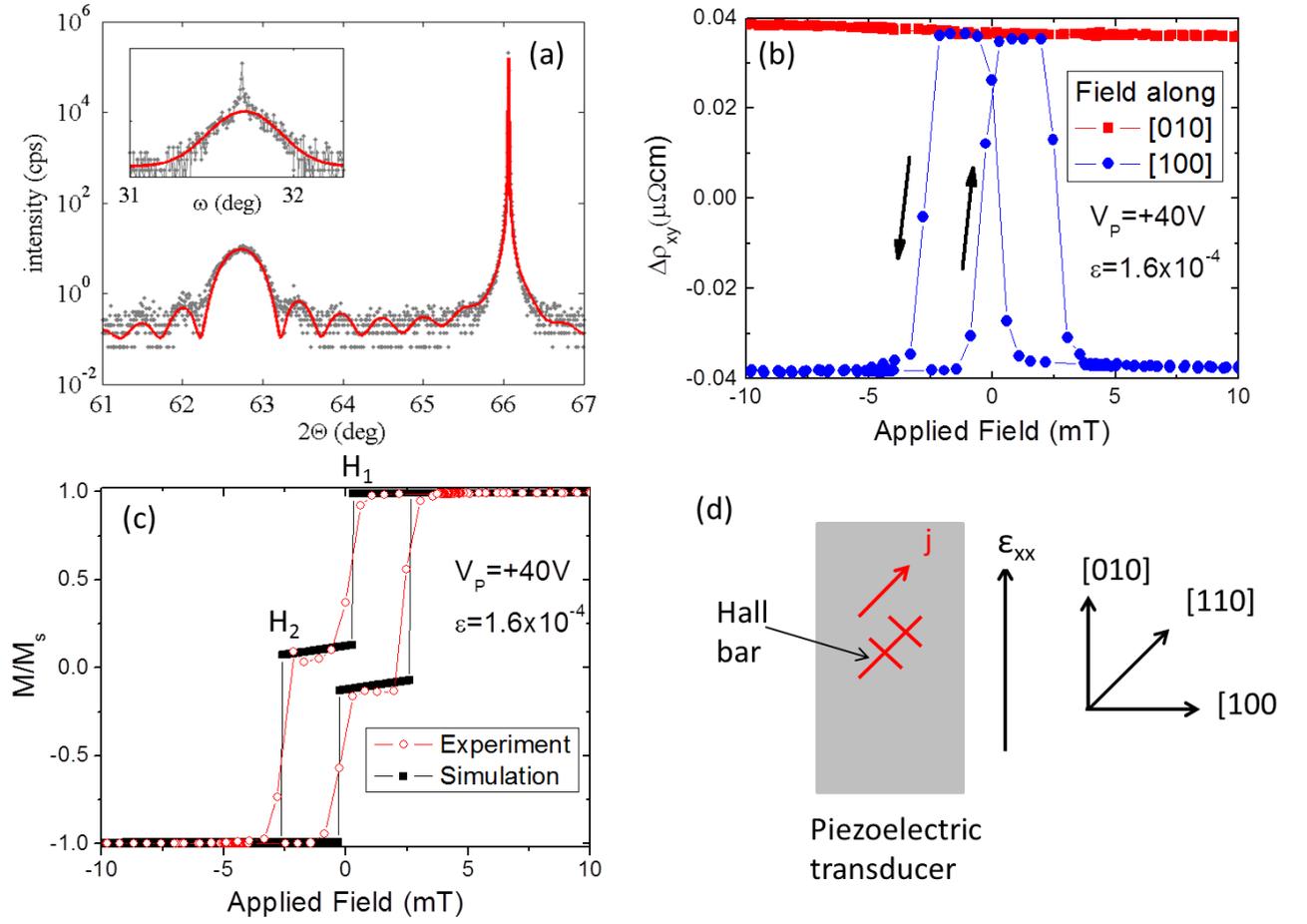

**Figure 1:** (a) X-ray 2θ/ω scan (main graph) and the ω scan (inset) of the MBE-grown thin $Fe_{81}Ga_{19}$ film. The points and lines denote the measured data and the fits, respectively. (b) The change in the transverse resistivity measured as a function of the magnetic field applied in the plane of the device along the [100]/[010] directions with tensile strain, $\varepsilon=\varepsilon_{xx}-\varepsilon_{yy}$ applied. (c) Magnetic hysteresis loop extracted from the data in (b) using the AMR formula for transverse resistivity and magnetic field applied along [100]. (d) Schematic diagram of the Hall bar/piezoelectric device layout showing directions of the tensile strain, $\varepsilon_{xx}$, the electrical current, j, and the crystalline directions of the $Fe_{81}Ga_{19}$ film.

Figure 1(b) shows the change in $\rho_{xy}$ measured at room temperature as an external magnetic field is applied in the film plane along the [100]/[010] directions for a positive voltage ($V_P$) applied to the piezoelectric transducer, corresponding to a tensile strain induced in the [010] direction in the Hall bar device. Large changes of $\rho_{xy}$ correspond to large changes of the magnetisation orientation as the external magnetic field is swept through zero, indicating that field sweeps along such directions correspond to sweeps along hard magnetic axes. Therefore, the magnetic easy axis can be identified as close to [010] for large tensile strain in that direction. The magnetic hysteresis curve for the field along the hard [100] axis, extracted from the resistivity data, is shown in figure 1(c). The shape of the magnetic hysteresis loop, involving a double step in each field direction, is characteristic of the magnetic reversal behaviour of a sample with biaxial anisotropy [24]. The switching fields labelled $H_1$ and $H_2$ correspond to the magnetisation switching between the [100]→[010] and [010] →[-100] directions respectively.

We modelled the hard axis magnetisation curves by minimising the magnetic free energy, given by:

$$E(\psi) = {K_C}/{4} \sin^2 2\psi + K_U \sin^2\left(\psi - \frac{\pi}{4}\right) - MH\cos(\delta - \psi) + E_{ME} \qquad (1)$$

where ψ is the angle between the magnetisation and the [010] crystal direction. The first term in this expression represents the cubic magnetocrystalline anisotropy energy density, favouring easy axes along the [100]/[010] directions. The second term is a uniaxial magnetocrystalline anisotropy energy density, typically found for ferromagnetic films deposited on GaAs(001) substrates [22], which favours the [110] direction for our films. Fits to superconducting quantum interference device (SQUID) magnetometry measurements, reported previously[25] for our films, yield $K_C$= 33kJm$^{-3}$ and $K_U$=9kJm$^{-3}$. The last two terms represent the Zeeman energy of the magnetisation M interacting with the external field, H applied at an angle δ to the [010] direction, and the magnetoelastic energy induced by the applied strain. This is given by:

$$E_{ME} = \frac{3}{2}\lambda_{100}(c_{12} - c_{11})(\varepsilon_{xx} - \varepsilon_{yy})\cos^2\psi \qquad (2)$$

where $\lambda_{100}$ is the magnetostriction constant, $c_{12}$ and $c_{11}$ are the elastic constants, and $\varepsilon_{xx}$ and $\varepsilon_{yy}$ are the relevant components of the strain tensor. The method for extracting the components of the strain tensor from our electrical measurements has been described previously[25]. It is not possible to separately extract the magnetostriction and elastic constants from our measurements due to the clamping of the thin film to the relatively thick substrate. However, for the purposes of magnetoelectric applications, the magnetoelastic constant $B_1 = \frac{3}{2}\lambda_{100}(c_{12} - c_{11})$ is the parameter of most interest. The model was fitted to the switching fields from the data obtained for a range of induced strains, by minimising the energy density in equation (1) with the additional condition that the magnetisation switching events between local energy minima occur when the reduction in the magnetic energy density equals the energy density required to de-pin a magnetic domain wall ($E_{DW}$). For this fitting $B_1$ and $E_{DW}$ were the only free parameters.

Figures 2(a) and (b) show $B_1(\varepsilon_{xx} - \varepsilon_{yy})$ and $E_{DW}$ as a function of strain, extracted for a range of voltages applied to the piezoelectric transducer. Also shown are the values that would be obtained for quench cooled alloys, taking values of 3/2$\lambda_{100}$=3.95x10$^{-4}$ and ($c_{12}$-$c_{11}$)=39.4GPa from ref.[15]. The magnetostrictive response of our thin film is as large as the values for the best bulk single crystals. This is likely due to the excellent crystal quality of our epitaxial $Fe_{81}Ga_{19}$ films. The diminished magnetostriction reported in polycrystalline thin films [16] arises from averaging of the crystal directions and may also be affected by grain boundaries which provide a mechanism by which the induced strain can relax.

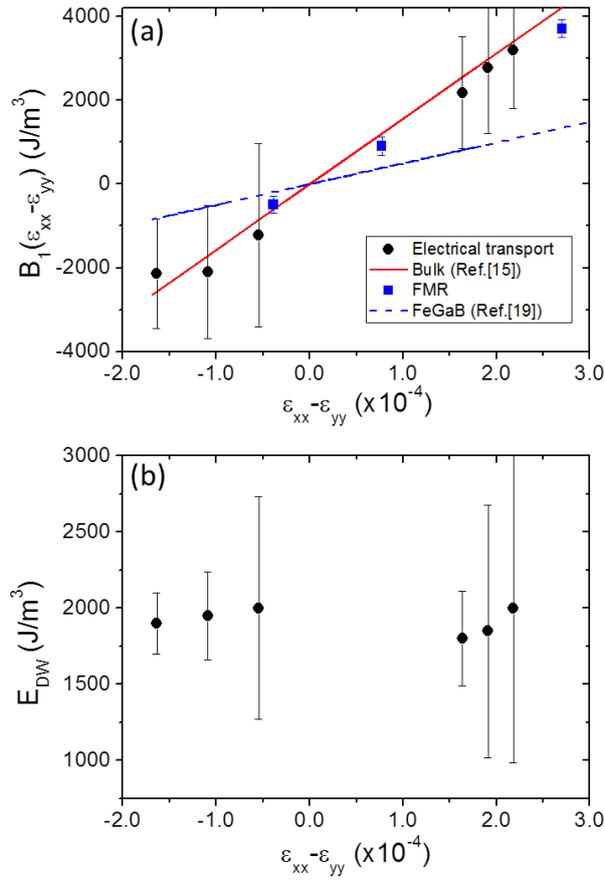

**Figure 2:** (a) The magnetic free energy coefficient induced by the strain for the single crystal thin $Fe_{81}Ga_{19}$ films obtained from electrical transport (black circles) and FMR measurements (blue squares). For comparison the curves deduced from the numbers reported for bulk single crystal (solid red line)[15] and thin FeGaB (dashed blue line) films[19] are also shown. (b) The domain wall pinning energy extracted from fitting to the electrical transport measurements.

The domain wall pinning energy, $E_{DW}$ is found to have no significant dependence on the induced strain. This can be understood if we consider pinning sites to occur where there are spatial variations in the quantity $E_{[100]-[010]}$ = E([100]) - E([010]), the difference in the anisotropy energy for the magnetisation along the two cubic directions. Homogeneous variations of the anisotropy energy due to the induced strain will not change the relative spatial fluctuations in $E_{[100]-[010]}$ and so will not alter the domain wall pinning energy. This observation will be important in designing and understanding the operation of devices which utilise strain as a method to control magnetic domain walls.

Ferromagnetic Resonance studies.

In order to measure the anisotropic FMR we utilized vector network analysis in combination with an octupole electromagnet capable of applying a field at any point in a sphere of 0.5T radius. Figure 3(a) shows a 2D resonance map obtained by measuring the s-wave parameter, $S_{12}$ which is related to the microwave transmission, as a function of frequency and applied field.

An expression for FMR is given by Smit and Beljers [26]

$$\left(\frac{\omega}{\gamma}\right)^2 = \frac{1}{M^2 \sin^2\theta}\left[\frac{\partial^2 F}{\partial \theta^2}\frac{\partial^2 F}{\partial \varphi^2} - \left(\frac{\partial^2 F}{\partial \theta \partial \varphi}\right)^2\right] \qquad (3)$$

where M is the saturation magnetisation, ω the resonant angular frequency, γ the gyromagnetic ratio and ϕ and θ are the polar and in-plane azimuthal angles of the magnetisation, respectively, which can be calculated with the equilibrium conditions,

$$\frac{\partial F}{\partial \theta} = 0, \frac{\partial F}{\partial \varphi} = 0, \qquad (4)$$

The magnetic free energy density, F, is an expression similar to Eq. (1), with additional terms to account for a magnetic anisotropy term and demagnetising field, favouring magnetisation lying in the plane of the film. By solving Eq. (3) under the conditions of Eq. (4), theoretical curves for the resonant frequency as a function of applied field were obtained and are in excellent agreement with the experimental data (red line in Fig. 3(a)).

Line scans through the frequency vs field map reveal the linewidth of the FMR, which is related to the damping in the ferromagnetic material. Figure 3(b) shows a horizontal line scan through Fig. 3(a) at a frequency of 15 GHz. The raw data is fitted to an asymmetrical Lorentzian function, which takes into account the coupling between the magnetic sample and the coplanar waveguide (CPW) which can partly mix the real and imaginary parts of the magnetic susceptibility. A striking feature of this epitaxial thin film is the narrow linewidth. Such narrow linewidths are indicative of a high quality, weakly damped system. Previous measurements of the FMR linewidth of bulk and thin film $Fe_{1-x}Ga_x$ have shown linewidths at X-band excitation frequencies (~10 GHz) between 18 mT [17] and 70 mT [18] for Ga concentrations around x=20%. In our epitaxial $Fe_{81}Ga_{19}$ thin films the FMR linewidth at 10 GHz is 7.7±0.3 mT along the [100] and [010] easy axes and 6.7±0.3 mT along the [110] hard axes, which is a factor of 2.5 to 9 lower than previous values found in the literature. Narrow linewidths are important in microwave devices, for example, in maximizing the power output in spin torque oscillator devices, as well as increasing the range of dc currents where the oscillator will lock to a reference signal.

The effect of the voltage-induced strain on the value of the resonant field along the [100] direction was investigated at 15GHz and is shown in figure 4(a). By applying voltages over the working range of the piezoelectric transducer in the hybrid structure we are able to tune the resonant field by 6mT. Such a property has significant implications for electric-field tunable magnetic microwave oscillators developed from such materials. The magnetoelastic energy can be extracted from figure 4(a) using the relationship:

$$B_1(\varepsilon_{xx} - \varepsilon_{yy}) = \frac{M \Delta H_R}{2} \qquad (5)$$

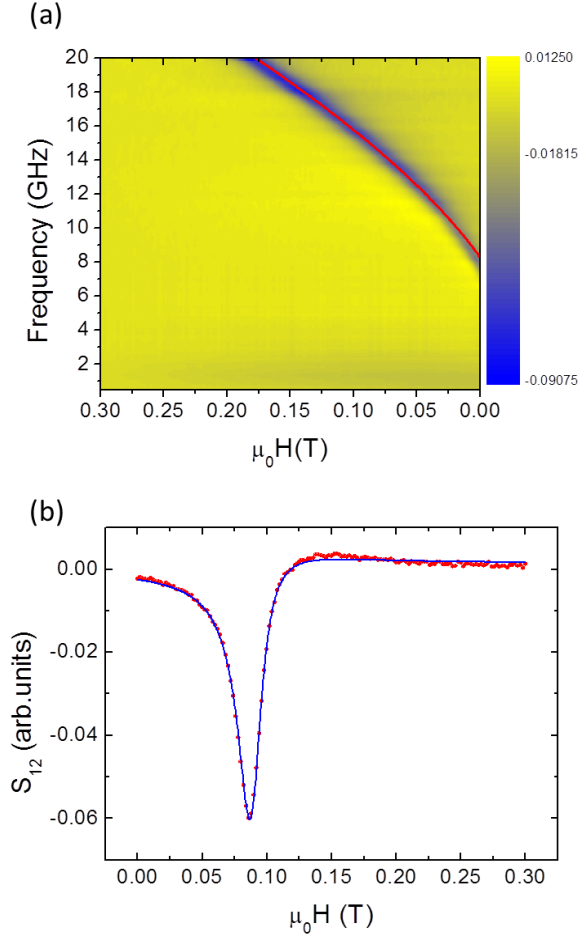

**Figure 3:** (a) Frequency vs magnetic field map of absorbed microwave power (VNA $S_{21}$) for the magnetic field applied along the [100] easy axis. The Kittel resonance can be clearly identified. The red line represents the theoretical curve. (b) The microwave transmission vs magnetic field measured from a line scan through (a) at a frequency of 15GHz. The data (red dots) are fitted to an asymmetric Lorentzian function (blue line) allowing the linewidth to be extracted.

where $\Delta H_R$ is the change in the resonance field in response to the induced strain. The values of $B_1(\varepsilon_{xx} - \varepsilon_{yy})$ extracted from the FMR measurements, shown in Fig. 2(a), are consistent with the large magnetostrictive response observed in the electrical transport device. Recent work by Lou et al [19] showed a giant tuning of the resonant frequency in the alloy FeGaB grown on the ferroelectric substrate PZN-PT. Using the values from ref. [19] ($\lambda_s$=60ppm, Young's modulus=55GPa), the corresponding curve for their FeGaB device is plotted in Fig. 2(a). The sensitivity of $B_1(\varepsilon_{xx} - \varepsilon_{yy})$ and of the resonant field to an applied strain is approximately three times larger in our $Fe_{81}Ga_{19}$ thin film than the films studied in ref. [19].

The effect of the voltage-induced strain on the linewidth of the FMR is shown in Fig. 4(b). The frequency dependent linewidth, *ΔH(f)* is related to the damping parameter, α by [27]:

$$\Delta H(f) = \Delta H_0 + \frac{2\pi \alpha f}{\gamma} \qquad (6)$$

From linear fits to the plots in Fig. 4(b) using Eq. (6) we are able to extract both the intrinsic damping, α, and the extrinsic damping, $\Delta H_0$, as a function of strain. The gradient of ΔH vs frequency is approximately equal for all applied strains, implying that the damping parameter, α = 0.017, is independent of strain. A similar measurement of the related composite multiferroic FeGaB/PZN-PT by Lou [28] showed a large variation of α with the voltage induced strain. Given that ΔH is dependent on the voltage induced strain, but that the gradient is independent, the only remaining factor to account for the change in linewidth is the offset, $\Delta H_0$, which is related to the extrinsic damping. The inset to Fig. 4(b) plots $\Delta H_0$ as a function of applied strain. The extrinsic damping is enhanced for non-zero tensile and compressive strain. Extrinsic damping is predominantly caused by inhomogeneous variations in the magnetic properties of the film. One possible explanation for the voltage dependent extrinsic damping is that the voltage-induced strain may vary slightly across the film, thereby modifying the magnetic anisotropy inhomogeneously.

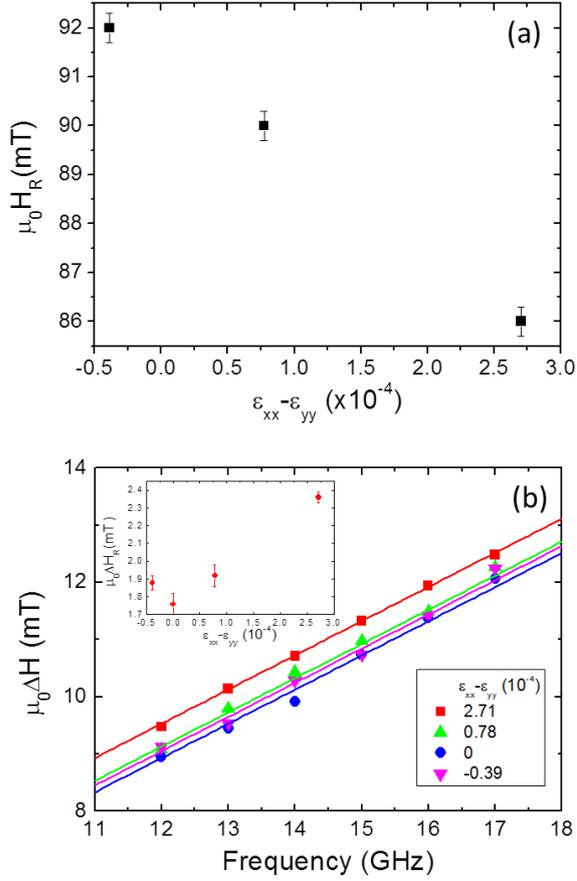

**Figure 4:** (a) The resonance field ($\mu_0 H_R$) along the [100] axis as a function of the strain at an RF frequency of 15GHz. (b) The FMR linewidth vs resonant frequency for different applied strain (points). Fits to the data using Eq. (6) are shown by solid lines. Inset: The intercept, $\Delta H_0$ as a function of the applied strain.

**Discussion**

Growth of $Fe_{81}Ga_{19}$ on the GaAs(001) surface by MBE results in high quality single crystal films. The films do not suffer from the reduction in magnetostriction observed in thin $Fe_{1-x}Ga_x$ films grown by other methods, and exhibit an inverse magnetostrictive response as high as the best bulk single crystal samples. The combination of large magnetostriction and cubic magnetic anisotropy has been shown previously to enable useful functionalities, including strain-mediated voltage-induced non-volatile switching of the magnetisation direction[25], and tunability of ordered magnetic domain patterns [29]. The demonstrations of such functionalities highlight single crystal $Fe_{81}Ga_{19}$ films as excellent candidates for applications in electric field controlled magnetic information storage[14] and logical processing devices[30].

We have shown that the large magnetostriction also enables large voltage-tunable shifts of the microwave resonant field and frequency. Compared to previous work on tuning of the microwave properties of Galfenol based thin films by a piezoelectric substrate [19, 28], our material achieves the largest reported change in effective field per unit strain. In addition, we find that the damping in the film, as measured around the X-band frequency, is substantially lower than previous reported values for $Fe_{1-x}Ga_x$ and is comparable to the amorphous FeGaB compounds. However, in the case of FeGaB, the addition of boron to reduce the linewidth is responsible for the diminished magnetostrictive effect. The combination of large magnetostriction and low damping in our single crystal thin films is very promising for applications in microwave frequency communications devices and concepts for logical processing using spinwaves.

The modifications of the resonance linewidth with applied strain are due to an increase in the extrinsic damping, likely caused by inhomogeneous strain, while the damping parameter, α remains constant. By scaling down the device to micron or submicron size, such inhomogeneities should be negated, reducing the extrinsic damping parameter further.

**Methods**

The 21nm molecular beam epitaxy (MBE) grown film was deposited by co-evaporation from Fe and Ga Knudsen cells at a substrate temperature of 0°C. The GaAs(001) substrate was first prepared by a high temperature (550°C) bake in vacuum followed by the deposition of a high temperature GaAs buffer layer. A 5nm amorphous GaAs capping layer was grown to protect the metallic layer from oxidation.

The crystal structure of the layers was investigated by high-resolution x-ray diffraction. We used a PaNalytical MRD diffractometer equipped with standard Cu x-ray tube, a parabolic x-ray mirror and a 4x220 Ge Bartels monochromator, as well as a 3x220 Ge channel-cut crystal analyser and a point detector. We measured a symmetric 2θ/ω scan crossing the 002 $Fe_{81}Ga_{19}$ and 004 GaAs maxima, and a ω scan across the 002 $Fe_{81}Ga_{19}$ layer peak.

For electrical transport studies, standard photolithography techniques were used to fabricate a Hall bar of width 45 μm with voltage probes separated by 235 μm, with the direction of the current along the [110] crystal direction (Fig.1(d)). Following a similar technique to Ref. [7] the chip was bonded onto a piezoelectric transducer capable of producing a uniaxial strain in the layer of order a few $10^{-4}$ at room temperature for applied voltages in the range -30V to +50 V. Uniaxial strain was induced along the [010] crystal direction with tensile strain defined as positive along this direction.

For FMR measurements, an unpatterned 5mm x 5mm sample, bonded onto a piezoelectric transducer, was mounted face down onto a 50Ω coplanar waveguide (CPW) connected to a two port vector network analyser (VNA) and centred between the poles of an octupole vector magnet. By measuring the microwave transmission losses whilst sweeping the microwave frequency as a function of bias field and angle, angular dependent FMR spectra were collected. In order to remove artefacts due to the CPW and cables a background trace, obtained by applying a field large enough to push the FMR above the frequency range of the VNA, was subtracted from each frequency scan. VNA-FMR spectroscopy over the microwave frequency range 0.5-20GHz was used to measure the effect of the voltage induced strain on the resonant frequency, linewidth and anisotropy.

**Acknowledgements**

The authors acknowledge financial support from EPSRC grant number EP/H003487/1 and EU grant No. NAMASTE 214499. We are grateful for useful discussions with Dr Jan Zemen, Prof. Bryan Gallagher and Prof. Tomas Jungwirth.

**Author Contribritions**

D.E.P processed the devices and carried out the transport measurements, L.S performed the FMR measurements, P.W, V. H, and A.T.H carried out the X-ray diffraction measurements and analysis, R.P.C carried out growth of the MBE films, S.A.C and G.v.d.L performed the macrospin calculations, S.A.C devised the FMR experiments, AWR devised the transport measurements, all authors contributed to the writing of the manuscript, analysis of the data and interpretation.

**Competing Financial Interests**

The authors declare no competing financial interests.